\documentclass[sigconf]{acmart}
\AtBeginDocument{%
  }

\setcopyright{acmlicensed}
\copyrightyear{2018}
\acmYear{2018}
\acmDOI{XXXXXXX.XXXXXXX}
\acmConference[Under Review]{Make sure to enter the correct
  conference title from your rights confirmation email}{}{}
\acmISBN{978-1-4503-XXXX-X/2018/06}

\NewDocumentCommand{\pquote}{m o}{{\textit{``#1''}\IfValueT{#2}{\space(#2)}}}
\usepackage{subcaption}

\begin{document}

\title{TractorBeam: Personalized AI Sensemaking Support via Collaborative Machine Annotation}

\author{Sireesh Gururaja}
\email{sgururaj@andrew.cmu.edu}
\orcid{0009-0004-0309-3393}
\correspondingauthor
\affiliation{%
  \department{Language Technologies Institute}
  \institution{Carnegie Mellon University}
  \city{Pittsburgh}
  \state{Pennsylvania}
  \country{USA}
}

\author{Nikolas Martelaro}
\email{nikmart@cmu.edu}
\orcid{0000-0002-1824-0243}
\affiliation{%
  \department{Human Computer Interaction Institute}
  \institution{Carnegie Mellon University}
  \city{Pittsburgh}
  \state{Pennsylvania}
  \country{USA}
}

\author{Emma Strubell}
\email{strubell@cmu.edu}
\orcid{0000-0003-2798-0726}
\affiliation{%
  \department{Language Technologies Institute}
  \institution{Carnegie Mellon University}
  \city{Pittsburgh}
  \state{Pennsylvania}
  \country{USA}
}

\renewcommand{\shortauthors}{Gururaja et al.}

\begin{abstract}

Language model-based systems have become popular tools for sensemaking. Despite their capabilities, including expanding the scope of sensemaking to whole corpora, these systems still suffer from issues of factuality and provenance, while encouraging primarily top-down sensemaking. We present TractorBeam, a browser extension-based mixed-initiative system that uses collaborative annotation as an interface metaphor for sensemaking, re-framing language model (LM) outputs as suggested highlights in a process that we call \textit{collaborative machine annotation}. This metaphor allows us to present LM results in-context on PDF documents, directly addressing concerns of provenance and factuality, while allowing users to iteratively construct mental schemata and queries for language models directly in the context of a document. Participants in our think-aloud and diary studies found that TractorBeam's collaborative annotation interface allowed them to iteratively construct useful, personalized highlighting systems, encouraged them to actively schematize their reading, and accomodated both top-down and bottom-up sensemaking.  

\end{abstract}

\ccsdesc[500]{Human-centered computing~Interactive systems and tools}
\ccsdesc[300]{Human-centered computing~Web-based interaction}
\ccsdesc[300]{Computing methodologies~Information extraction}

\keywords{Sensemaking, Collaborative Annotation, AI Reading Support, Personalized Reading Support}
\begin{teaserfigure}
    \centering
    \includegraphics[width=1\linewidth]{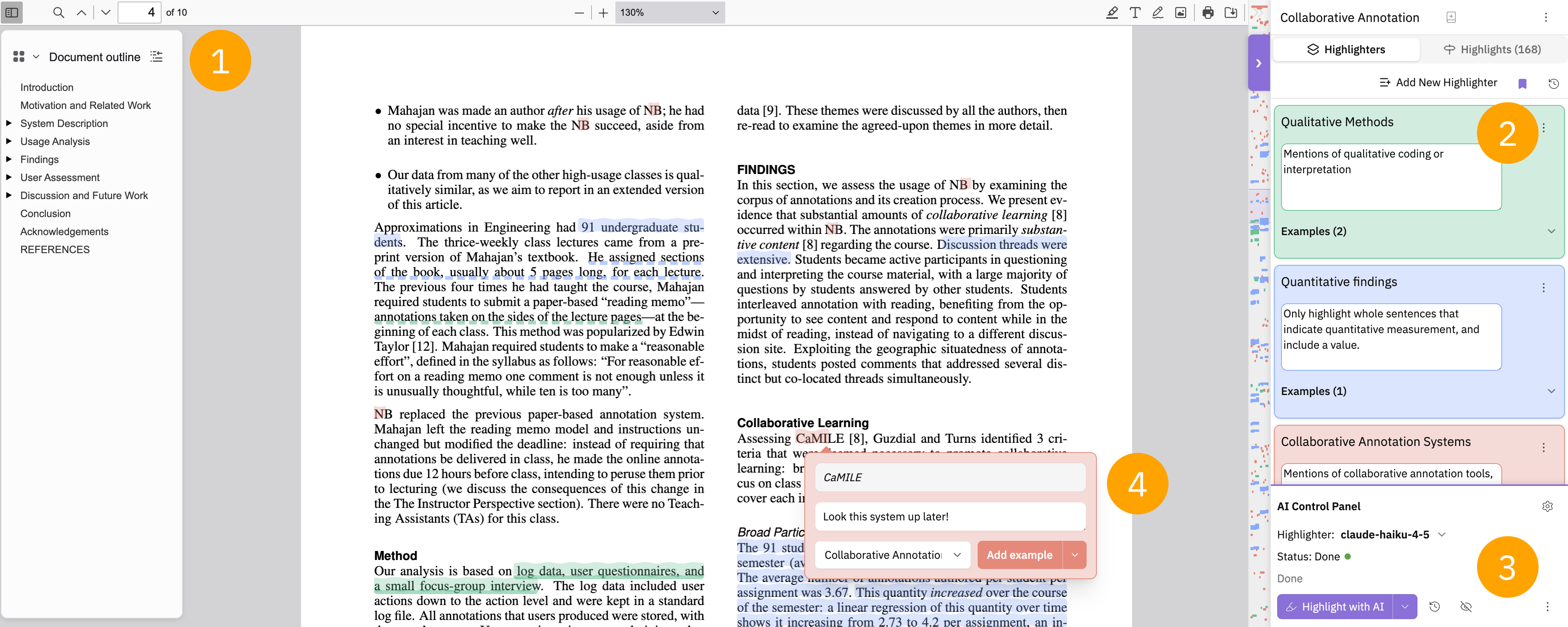}
    \caption{A screenshot of TractorBeam open on a paper, showing 1. The PDF viewer, where both human and LM annotations are displayed directly on the PDF; 2. The “highlighters” view, where users can specify their own schema for annotation; 3. The AI control panel, where users can configure and trigger AI highlighting; 4. An example modal dialog for users to accept or reject LM highlights before iteratively re-annotating. In the viewer, solid and dashed underlines indicate user-provided positive and negative examples, and squiggly highlights indicate AI highlights. The PDF in this example is \citet{zyto_successful_2012}}
    \label{fig:teaser}
\end{teaserfigure}

\received{10 September 2026}

\maketitle  

\section{Introduction}
\label{sec:intro}

Making sense of large document collections is a fundamental task for many kinds of knowledge work. Though the specifics and goals may vary, the need to construct coherent understanding of a corpus of documents persists across fields and domains. As a result of this need, technological assistance for sensemaking has a long history in the human-computer interaction literature \citep[][\textit{inter alia}]{russell_cost_1993, m.russell_sensemaking_2026}. The foci and affordances of sensemaking systems over this period have varied widely, from tools that visualize document collections in various ways, for example facilitating iterative or collaborative search for documents \citep{baldonado_sensemaker_1997, paul_cosense_2009}, visualization of documents based on document attributes, whether from metadata, automatically inferred, or user-specified \citep{stasko_jigsaw_2008, hePaperPolesFacilitatingAdaptive2019, chang_searchlens_2019}, or in-context annotation of document content \citep{fok_scim_2023, august_paper_2023}.


 More recently, large language models (LLMs) have increasingly been pitched as solutions that expand the scope of sensemaking. LLMs allow users to express their intent to a deployed system via natural language queries, and enable more user-specific processing of documents across entire corpora. For example, systems like ScholarQA~\citep{asai_synthesizing_2026} and DocWrangler~\citep{shankar_steering_2025} facilitate complex processing over collections of documents, and LM-based assistance for qualitative coding is an active area of research~\citep[][inter alia]{gao_collabcoder_2024, marathe_semiautomated_2018, schroeder_large_2025}. However, though these systems allow users to specify complex and specific intents, they nonetheless raise a number of concerns.

 First, the output of these systems often takes the form of generated text. Validating the provenance and accuracy of the information in generated text remains a challenge. While citations have often been presented as a fix for this problem, \citet{ding_citations_2025} found that the presence of citations can increase trust in results even as following those citations decreases it; \citet{onweller_cited_2026a} find that even where citations are referentially correct, the cited sources often do not support the evidence presented. Earlier sensemaking systems like Scim \citep{fokScimIntelligentSkimming2023} or PaperPlain \citep{august_paper_2023}, though statically defined, do not suffer the provenance issue by displaying their suggestions as in-context augmentation of documents, which makes evaluating accuracy and provenance much simpler.

 More generally, LM-based sensemaking tools typically present the LLM---whether as a chat interface or otherwise---as the primary interface to the process of sensemaking. This results in what \citet{pirolli_sensemaking_} describe as a "top-down" process: sensemaking begins in the LLM output rather than in the data itself. In the case of chat-based systems like ScholarQA, requires the user having to have a well-defined question to answer before using the system; in the case of systems like DocWrangler, this requires the user to have a set of desired results and a way to evaluate suggestions before engaging with the system at all. Sensemaking, however, is often iterative, moving from data to schema ("bottom-up sensemaking") in addition to the top-down version that these tools facilitate in a mutually constitutive relationship.

LLM-based systems are also pose challenges to users because of the unpredictability of LLM performance even on similar tasks, the so-called “jagged frontier” of LLM performance \citep{dellacqua_navigating_2023}. Worsening this further is the degree to which LLM performance can depend on specific prompting patterns that users may be unaware of \citep{zamfirescu-pereira_why_2023}. Prior work has studied extensively how to scaffold user-driven , iterative development of LLM systems on their own terms\citep{romerolauro_rag_2026, arawjo_chainforge_2024}, and we argue that those tools must also be deployed in the service of sensemaking. 

To better support LM-assisted, iterative sensemaking, we present TractorBeam, a browser extension-based system that conceptualizes sensemaking assistance as a process of collaborative annotation on document content between a user and an LLM, a process that we call \textit{collaborative machine annotation}. TractorBeam draws on the extensive study of collaborative annotation in educational settings \citep{davis_shared_1995, cadiz_using_2000, hwang_study_2007, su_web_2010, zyto_successful_2012, miller_analysis_2016, goller_sharing_2021} that establish that annotation, both collaborative and individual, can increase engagement in reading. The placement of annotations directly on the originating documents encourages bottom-up sensemaking as users read, while at the same time offering them the ability to engage with documents top-down through annotations as suggested points of interest. Users can additionally interactively evaluate and edit these annotations, as in systems like PaTAT \citep{gebreegziabher_patat_2023}, ChainForge~\citep{arawjo_chainforge_2024}, or Raggy\citep{romerolauro_rag_2026}. As users develop deep familiarity with the models they build, these models can also be extended to support top-down, corpus-level sensemaking as in systems like DocWrangler\citep{shankar_steering_2025}.

\section{Related Work}

\paragraph{AI-assisted Sensemaking} We attempt to situate TractorBeam in between two closely related lines of work. First, earlier reading assistance-based sensemaking tools that render their suggestions in-context on the document, like Scim \citep{fok_scim_2023} and PaperPlain \citep{august_paper_2023}. Second, TractorBeam depends on the flexibility and expressive power of LLMs, as exemplified by sensemaking systems like ScholarQA \citep{asai_synthesizing_2026} and DocWrangler \citep{shankar_steering_2025}. TractorBeam aims to provide users the affordances of in-document highlights, while allowing them the flexibility to determine what should be highlighted. 

\paragraph{Collaborative Annotation} TractorBeam's design frames LLM sensemaking assistance as collaborative annotation on a PDF document between a reader and the LLM. We design around collaborative annotation to draw upon its well-documented benefits to reading, both in individual and collaborative settings. \citet{hwang_study_2007, su_web_2010, miller_use_2018} demonstrate, for instance, that collaborative annotation systems help students' performance on evaluations in both individual and collaborative settings. \citet{zyto_successful_2012} additionally discuss the benefits of highlighting specifically as it pertains to the spatial dimension: highlights serve as an index of the content of readings, to direct students' attention. TractorBeam operationalizes this directly with its minimap feature, as outlined in section \ref{sec:design}.

\paragraph{Qualitative Coding and Information Extraction} Highlighting document content is also closely related to the processes of qualitative coding as part of methods like thematic analysis and grounded theory, and to information extraction in natural language processing. Work in qualitative coding has studied both the formation of the codes, as well as their application with AI; \citet{jiang_supporting_2021} find that researchers wish to retain agency over the creation of the codes, but both they and \citet{marathe_semiautomated_2018} find that qualitative researchers would like support in the application of codes to new documents. TractorBeam's design is influenced by best practices in qualitative coding assistance in centering the user's role in defining what should be highlighted, while at the same time allowing an evolving set of “codes” to be applied iteratively to both new and old documents. We also derive inspiration from the vision of next-generation information extraction outlined in \citet{katz_neretrieve_2023}, which envisions a process of information extraction that is not merely a search for predefined entities, as in classical named entity recognitions, but a dynamic process that centers user-defined intent to allow for the detection of fine-grained, idiosyncratic categories of entities such as dinosaurs and dog breeds.

\paragraph{Scaffolded End-user Programming} The advent of LLMs has seen a rise in the degree to which end-users are expected to specify the behavior of computer systems. As outlined in section \ref{sec:intro}, while LLMs allow for granular specification of user intent, reasoning about the effects of those changes can be challenging. Prior work handles this in different ways that nonetheless focus on allowing users the ability to confidently make changes in LLM systems and reason about the results. ChainForge \citep{arawjo_chainforge_2024}, for example, facilitates the comparison and evaluation of multiple LLMs by integrating evaluation results directly into the prompt development flow. PaTAT \citep{gebreegziabher_patat_2023} takes a different tactic, transforming user annotations of qualitative codes into interpretable combinations of lexical and grammatical rules that are then used in place of an LLM. PaTAT additionally allows users a great deal of flexibility in managing their codes through renaming, merge/split operations, or outright deletions. TractorBeam aims to scaffold the user specification of LLM behavior with both direct evaluation of annotations on PDF content, as well as a similar flexible annotation management system to PaTAT. 

\section{Background and Design Goals}
\label{sec:background+design}

TractorBeam builds on three lines of work to inform its design goals. First, work that characterizes expert reading processes \citep{gururaja_text_2025, bazermanPhysicistsReadingPhysics1985, mysoreHowDataScientists2023}, to inform where AI can support existing experts' process of reading. Second, the closely related literature on AI support for qualitiative coding \citep{jiang_supporting_2021, marathe_semiautomated_2018, dunivin_scaling_2025}, which both identifies some similar needs to the reading support literature, as well as suggesting best practices for AI support. Finally, we incorporate the ML Ops perspective from \citep{shankar_operationalizing_2022}. We discuss how each of these threads informs TractorBeam's design below. 

\citet{gururaja_text_2025} study the reading processes of experts across two domains, and find similarly to \citet{bazermanPhysicistsReadingPhysics1985, mysoreHowDataScientists2023} that experts' reading processes are deeply idiosyncratic, and typically involve the continual, iterative development of mental schemata as reading progresses. They outline four principles for AI-supported reading: that systems should allow users to personalize systems to their own needs, rather than only those foreseen by the developer; that that personalization must allow for iteration and development as the goals for the reading support system, or the reading itself, shift; that the tools must support users in reasoning about the social context of documents, and how that social context informs perceptions of relevance and quality; and finally, that these tools must be accessible, in that they should allow users to define their reading support without the burden of context switching to a dedicated annotation tool, or to code.

The iterative construction of idiosyncratic frames mirrors findings from studies of qualitative coding, which also relies on the affordances of highlights, typically as a preliminary step. \citet{jiang_supporting_2021}, for instance, describe a common process across qualitative traditions of finding “data snippets of interest” before iteratively attempting to categorize or describe them. In addition to the shared focus on idiosyncrasy and iterativity, they additionally describe their participants focus on uncertainty, ambiguity, and serendipity: participants feel adamantly that they do not wish for the automation of this type of analysis, and that the messy processes of finding passages and categorization constitute the sensemaking process, rather than being barriers to it. Later work in AI-supported qualitative work has reified this focus: \citet{than_updating_2025}, for instance, centers the iterative loop as part of developing a codebook in their study of coding a news corpus; \citet{ibrahim_should_2026} additionally suggests documenting the iterative development of AI instructions and codebooks is analogous to recording interactions with research participants. Unlike in qualitative coding, however, we expect categorization for reading support tools to remain incomplete as users continue to read.

Iterativity is not always simple to enable, given the unpredictability of LLM behavior. Small user changes in the wording of prompts, for example, can cause unpredictable changes in a model's predictive performance. Even were the user to configure, prompt, and otherwise set the system up perfectly, the underlying models themselves have limitations---social reasoning, for example, is an understudied aspect of model behavior, with early results indicating that models are often far too credulous \citep{yun_caught_2025a} and make unpredictable inferences from seemingly unrelated demographic and cultural cues \citep{li_chatgpt_2024, kantharuban_stereotype_2025}.  \citet{shankar_operationalizing_2022} outlines a three-principle framework for machine learning operations (MLOps) that characterizes how industry practitioners effectively build and operationalize machine learning-based systems over time: they focus on the “three Vs” of velocity, validation, and versioning. Velocity is primarily important due to the unpredictability and experimental nature of machine learning systems: being able to try several modeling choices easily in order to determine which might work enables effective development. Validation refers to the (typically quantitative) ways to assess whether the impacts of a change are desirable or not, and versioning refers to the ability to roll back to a previous state known to be good if the experimentation is failing to produce a model that meets desired performance characteristics. 

We see several synergies between these three sets of goals---reading assistance, qualitative coding, and MLOps---that inform the following design objectives for TractorBeam:

\textbf{Personalization amid Uncertainty} The primary purpose of the system should be to support readers’ idiosyncratic processes. Any support that the tool offers must reflect user priorities, and as such, should not attempt to anticipate user needs or otherwise act without explicit user interaction, so as not to coerce the user to resolve ambiguity and uncertainty before they are ready. The system should not require user input before the user has an information need---it should support users organically coming to information needs as they read. 

\textbf{Iterativity} Users must be able to modify, reconfigure, or otherwise change the functioning of the tool, in order to tailor the support provided by the tool to their needs. In addition to merely \textit{allowing} this iteration, however, the tool should proactively scaffold this iteration such that users can effectively reason about changing the functioning of the tool, in keeping with the MLOps principles above: changes must be easy, viewing and evaluating the results of those changes should be fast, and users must be able to roll back to a previous version of the system if the changes are ultimately not useful.

\textbf{Design around failure} In addition to the iterativity of the users' reading, best practices for machine-learning based systems also hinge on iterativity, because of the relatively high likelihood that the change a user makes to a system will have unpredictable consequences. As a result, the system must design around the potential failures of models to accurately reflect the user's intent, whether because of underspecification or fundamental limits. Further, the tool should allow \textit{differential use}---users should be able to use the provided support to the degree that it is useful, without that support being an all-or-nothing choice. 

\textbf{Accessibility} Finally, the tool should aim to minimize the degree of context switching required of users. More traditional approaches to the development of ML approaches involve pipelines spread across multiple bespoke tools, often used by people in different roles: computational notebooks for data scientists conducting exploratory analysis, annotation tools with data in custom formats for annotation and resolving disagreement, and separate tools still for training the model, evaluating its predictions, and monitoring its performance. These choices reflect the separate expertise and effort that go into maintaining machine learning systems. Given that this tool aims to allow individual users to develop reading assistance for themselves, this staged pipeline approach becomes impractical. To the extent possible, the exploration, analysis, training, and evaluation should happen inside the user's existing process, so as to minimize the effort required of a user to keep their reading support models up to date.  

\section{TractorBeam}
\label{sec:design}

In this section, we present TractorBeam, the system that we designed and implemented in response to the design goals of the previous section. TractorBeam is a browser-extension based system that allows users to create, evaluate, and iterate on personalized AI sensemaking systems from within the context of PDF documents. TractorBeam frames this AI support as the output of a collaborative annotation process, where both user highlights and AI outputs are visualized as highlights on document content. We term this process \textit{collaborative machine annotation}. In this section, we discuss the design choices that we made and how they were shaped by our goals. 

\subsection{Collaborative Annotation in the PDF Viewer}


\begin{figure*}
    \centering
    \includegraphics[width=0.85\linewidth]{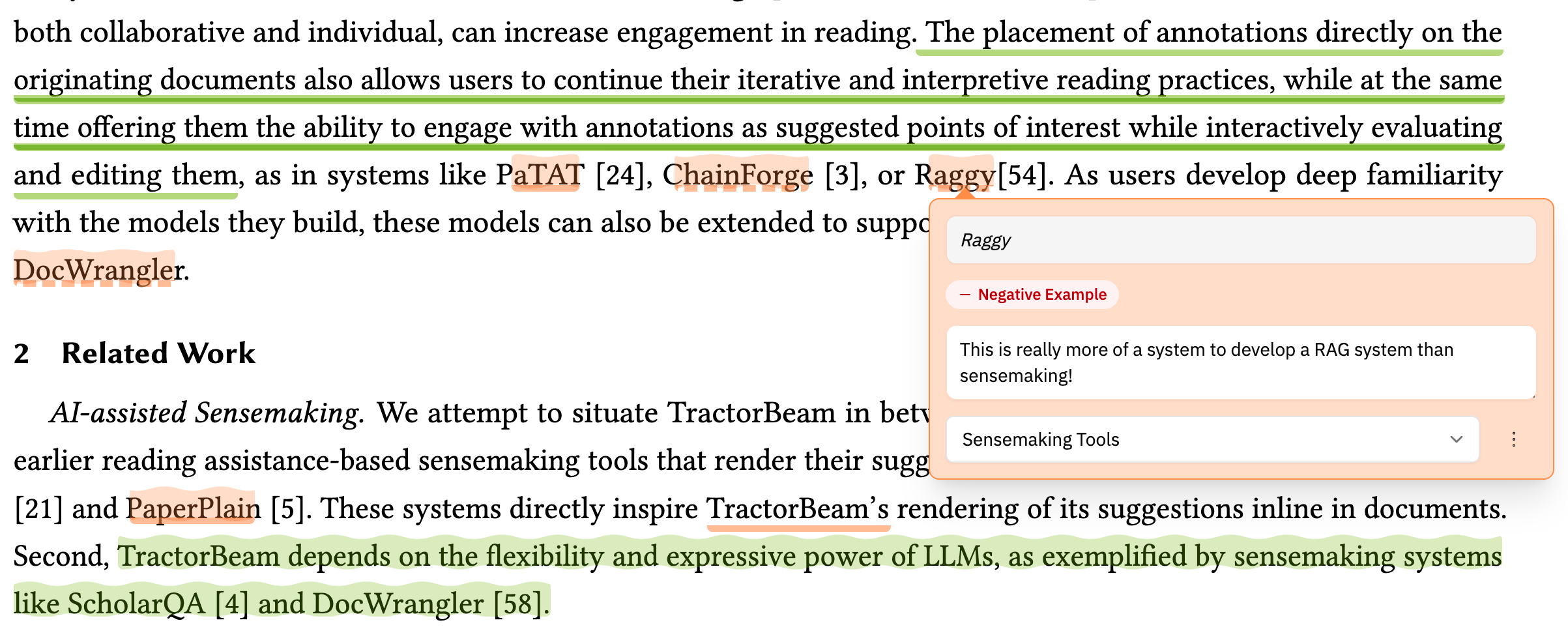}
    \caption{A closeup of the different types of annotations in TractorBeam. User annotations are displayed as underlines to indicate certainty, while machine annotations are squiggly to indicate uncertainty. Dotted underlines indicate a negative example, which users typically use for model corrections. Human and machine annotations can be seen simultaneously, indicating where models have correctly recovered human annotations.}
    \label{fig:annotations}
\end{figure*}

Users access TractorBeam by launching the extension when they are viewing a PDF document. TractorBeam then opens in its default view, as seen in Figure \ref{fig:teaser}: a PDF viewer and a sidebar containing the controls for the user and machine annotation (which we discuss below). We chose to extend the browser and present a largely unmodified PDF reading experience to address our goal of accessibility: PDFs are often read in-browser, and staying in this environment means less disruption to existing workflows. TractorBeam can additionally access local files, and is therefore compatible with the workflows of users that maintain local PDF libraries with systems like Zotero. 

Similarly, we choose to vendor in a copy of Mozilla’s pdf.js viewer---the default in the Firefox Browser---rather than implement a new PDF viewer, as in AI2’s Semantic Reader project \citep{lo_semantic_2024}. This provides users with an instantly familiar PDF experience that simultaneously frees us from re-implementing common PDF functionality. The centrality of the PDF viewer to this implementation also underlines how reading is the primary modality for this tool---users, upon first opening TractorBeam, can begin by reading for bottom-up sensemaking, though machine annotation is available if they so choose. This supports differential use: should the user not wish to engage in annotation at all, the controls sidebar can be minimized, and the user's experience gracefully degrades to a familiar PDF viewer.

In the PDF viewer, we present a similar user experience as presented by existing collaborative annotation tools like Hypothesis\footnote{\url{https://hypothes.is}} or Margin\footnote{\url{https://margin.at}}. Users highlight text in order to annotate it, and can additionally write notes that are stored alongside highlights. In addition to this, we present the user with visual indications that indicate provenance and epistemic status, as seen in Figure \ref{fig:annotations}: user and machine annotations are visualized differently. User annotations are indicated as underlines, and machine annotations are indicated as squiggly highlights. Their relative visual weight corresponds to how much we intend for them to call attention to themselves: users are presumably already aware of what they have annotated, requiring less visual weight; machine annotations are intended to direct attention. 

Collaborative annotation presents a number of properties that we believe effectively support the definition, validation, and iteration on desired LLM outputs. First, collaborative annotation on the part of the user is data annotation. Presenting data annotation in-place in a PDF viewer allows for a significantly more integrated experience for annotating data than common, standalone tools like INCEpTION \citep{klie_inception_2018} or brat \citep{stenetorp_brat_2012}: users now annotate and evaluate in-place in a tool that they might already use in their workflow. Further, while recent LLMs allow users to specify their behavior in terms of prompting, which can be extremely effective in the low-data regime \citep{lescao_how_2021, liu_fewshot_2022, bertsch_incontext_2025} (a feature that TractorBeam takes advantage of, see sections \ref{sec:schema_management} and \ref{sec:ml_annotation}), providing example instances can be a powerful complementary input, as seen with in-context learning \citep{brown_language_2020} or even fine-tuning, which allows for more granular representation of intent, and can also work on much smaller models for more efficient inference. Prior work on evaluating LLMs for codebook following \citep{fonseca_can_2024, halterman_codebook_2024} indicate that fine-tuning can often be necessary for LMs to label nuanced concepts.

Second, collaborative annotations are located both on and in document content: the annotation is not separated from the context that generated it, and is located close to annotations on nearby content.
This implicitly supports the validation of machine performance. Because LLM outputs are not abstracted away from the documents, understanding whether or not they satisfy the user intent is much simpler than following citation links. Users can also over time build a deep, qualitative understanding of a model’s performance, rather than relying on summative statistics that do not accurately convey the experience of using a model. Machine annotations' groundedness also preserves serendipity in the process of reading, which \citet{binst_what_2025} define as \textit{“user experience in which the user
unintentionally encounters content that feels fortuitous, refreshing,
and enriching.”} Regardless of the correctness of the highlights, users are able to look through the surrounding context to find content that they may not have known they wanted, allowing for the iterative, exploratory phase of reading to continue, even where their attention is being directed by machine annotations. 

\subsection{Managing User schemata}
\label{sec:schema_management}


\begin{figure}
    \centering
    \includegraphics[width=0.75\linewidth]{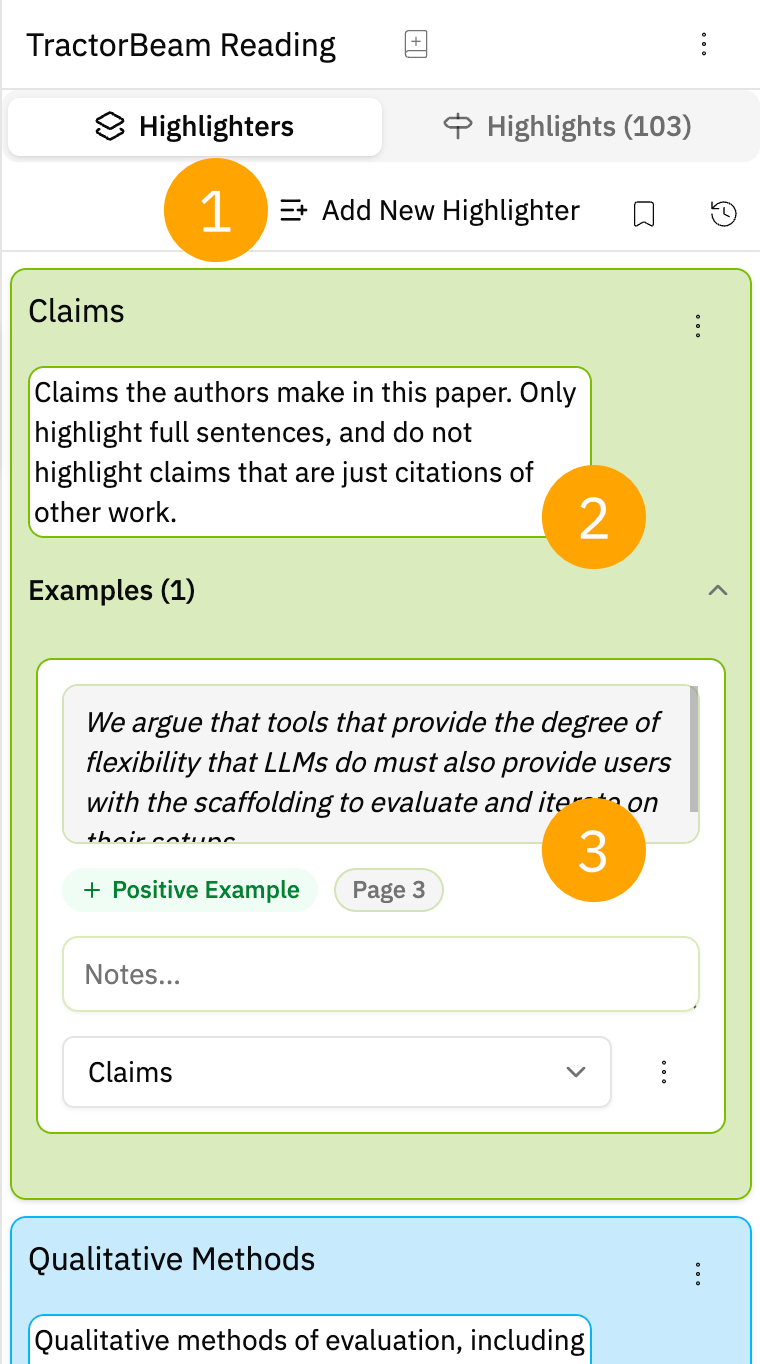}
    \caption{Closeup of schema viewer, showing: 1. Controls that allow for schema management, including adding new classes (expressed as “highlighters”), bookmarking specific schema versions, and rolling them back; 2. Schema classes themselves, which consist of a title, a description, and list of examples (all editable); and 3. an example of the schema class, i.e. a highlight on the current document. The notes are freely editable, the example can easily be reassigned to another class, and clicking on the example navigates to it in the document.}
    \label{fig:schema_manager}
\end{figure}

Annotation in TractorBeam is schema-driven. Schemata (which we term “projects” for users) are composed of a list of schema classes (termed “highlighters”), which indicate different types of things to be annotated. Highlights, whether user or machine, must belong to a schema class. Schema classes are defined by a name, optional description, and a list of instances; each schema class additionally has an associated color, which the highlights inherit. Schemata contain no classes by default: the construction of schemata, i.e. the goals for the reading support, are entirely user defined, enabling tailored annotation for exactly what the user is looking for. Users can maintain multiple parallel schemata for different kinds of annotation. Figure \ref{fig:schema_manager} shows a closeup of the TractorBeam schema editing view, with primary editing affordances highlighted.

TractorBeam’s schemata are designed to be highly flexible: the names of the schema, class names, and class descriptions are all freely editable at any time, new classes can be created or deleted, and instances can be freely re-assigned between categories or deleted. While we do not explicitly support merge and split operations, these are achievable as reassignments of instances. Aside from the constraint that highlights belong to a schema class, this imposes very little upfront structure: users are free to determine how much structure they would like at any point during the annotation process. A user could, for example begin with a catch-all “interesting things” class that contains all of their highlights, and create new classes as they begin to construct a mental model, reassigning highlights to the new classes that they create along the way. This supports the preservation of uncertainty, and the iterativity that comes in both reading and qualitative coding: classes can be created, progressively refined, or even completely redefined as the process of reading demands. Despite the freeform structure, however, we do not allow uncategorized annotations to encourage users to actively consider categorization and schematization while using the tool, in keeping with the model in \citet{pirolli_sensemaking_}.

\subsection{Machine Assisted Annotation}
\label{sec:ml_annotation}


 \begin{figure*}[htbp]
    \centering
    \begin{subfigure}[t]{0.32\textwidth}
      \centering
      \includegraphics[width=\linewidth]{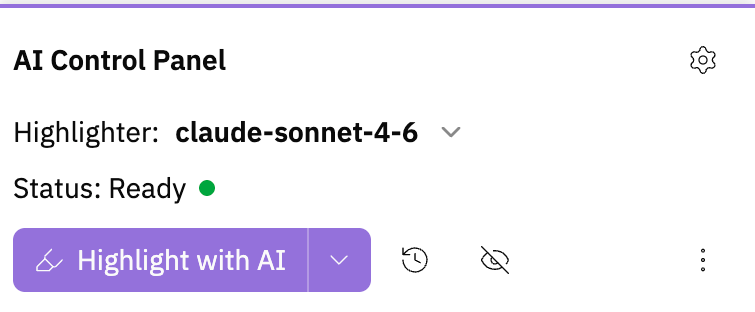}
      \label{fig:ai_controls_basic}
    \end{subfigure}
    \hspace{0.03\textwidth}
    \begin{subfigure}[t]{0.32\textwidth}
      \centering
      \includegraphics[width=\linewidth]{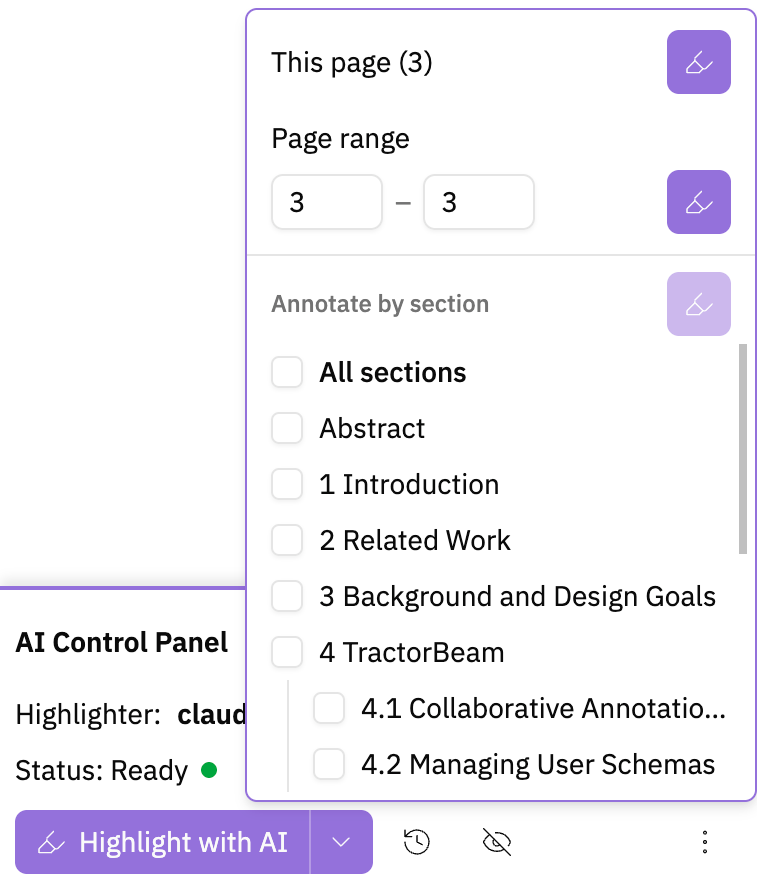}
      \label{fig:ai_controls_expanded}
    \end{subfigure}
    \caption{(a) Controls for machine annotation. Users trigger AI annotation by clicking the “highlight with AI” button. They choose the model to run against document text by selecting the caret next to the model name, and configure model backends via the gear icon. Previous annotation runs are viewable via the history button immediately next to the “highlight with AI” button, and annotations can be dismissed with the hide icon. (b) Annotation controls for finer-grained annotation, allowing users to apply machine annotation to a single page, page range, or document section.}
    \Description{Two views of the AI control panel, by which users control AI annotation. The panel displays the model the user has selected, its status, and controls for annotation.}
    \label{fig:ai_controls}
  \end{figure*}

LLM annotations in TractorBeam are controlled through the panel in the bottom right (see element 3 in Figure \ref{fig:teaser} and Figure \ref{fig:ai_controls}). As with requiring user definition of the schemata for support, TractorBeam requires users to trigger ML annotation, rather than it being automatically applied to documents. This is also designed to allow users to engage with the system in ways that suit where they are in the sensemaking loop: they can begin either by looking at the document itself, or with machine annotations to guide their attention. The machine annotation’s goals are entirely defined by the user’s schema. When the user clicks “highlight with AI”, TractorBeam constructs a system prompt that specifies what to highlight using the schema class names, descriptions, and examples. We use a popular form of LLM-based structured information extraction\citep{paolini_structured_2020} that frames text highlighting as a sequence-to-sequence translation problem, where the input sequence is the original text, and the output sequence is the original text augmented with XML tags indicating the entity type. This form of information extraction allows us to correctly calculate the span of the entity text in the original context, and find its position in the document deterministically. To keep the document content size appropriate for a wide range of LLM context windows, we use the geometric XYCut algorithm to parse document layout into visual chunks that typically align with paragraphs. This architecture assumes the use of a prompt-based LLM; we make this choice for the well-documented ability of such LLMs to take instructions and incorporate examples with some degree of success\citep{brown_language_2020}. That said, we implement the ML functionality behind an interface such that were we to use models with different affordances, e.g the GLiNER line \citep{zaratianaGLiNERGeneralistModel2023, zaratiana_gliner2_2025}, that change is not burdensome.

Within the tool itself, TractorBeam additionally allows for user configuration of what models will be run: we support both OpenAI- and Anthropic-compatible backends, and users are free to specify the base URL and bring their own API key to configure what service their reading support will rely on. The support for user supplied base URLs extends TractorBeam’s compatibility to popular third-party LLM hosting services, as well as self-hosting options like Ollama, vLLM, and SGLang. Users are additionally free to select from a list of available models within each provider. We provide this degree of flexibility to offer users control over the machine annotations applied to their documents, along multiple axes: what specific models are run on document content, where the document content is allowed to be stored (especially important in the case of e.g. internal company documents, or academic papers whose publishers disallow LLM use on their material), and the cost of inference. We argue that this degree of control is necessary for any tool that intends to see long-term use. In the use of machine learning systems, users often build mental models of how the ML models will function, and replacing that model with another model, even one that is considered stronger or demonstrates better evaluation results, can nonetheless result in qualitative changes that result in the invalidation of a user’s mental model \citep{bansal_updates_2019}. Over time, this can result in an erosion of trust in the LLM system \citep{damour_underspecification_2020, ma_schrodingers_2024}, a problem that is accentuated by LLM API providers' tendency to either make silent updates to models, or to deprecate popular models with little notice. In the spirit of user control, we additionally experimented with LLMs run entirely within the browser using the WebLLM project \citep{ruan_webllm_2026}, but found that the supported models’ latency and performance were ultimately insufficient to provide a reasonable user experience; we intend for the integration of in-browser models to be a line of future work.

In addition to using models to annotate documents, users can use the model highlights to further specify the behavior of those models. We present two affordances for this. Users can observe patterns in the observations and update the description of their schema classes to include more specific instructions (e.g. “Highlight whole sentences”). They can also directly indicate on a machine highlight whether it is correct or incorrect, by adding it to the relevant schema class as either a positive or negative example. These examples then take on the status of user annotations, and their positive/negative status is also indicated in the style of the underline (see Figure \ref{fig:annotations}. This type of feedback is available to users, but is never required: the purpose of this tool is to facilitate users developing their own reading assistance tools, rather than curating a comprehensive dataset; to require this type of feedback on each annotation would require enormous amounts of time and effort for unclear benefit. This is additionally why TractorBeam does not attempt to quantitatively evaluate its highlighters: the precision or recall of a highlighter is variably correlated to its downstream utility as a highlighting model, and we do not consider optimizing for those numbers a productive use of users’ time. 

Our approach to machine annotation relies extensively on models considered “black box”, whose internal function is ultimately poorly understood. In this, we make notably different choices from similar tools like PaTAT \citep{gebreegziabher_patat_2023}, which argues that interpretability is a necessary factor in building trust for machine annotations, and builds a system that infers interpretable lexical and grammatical rules from the user’s annotated instances. This approach prioritizes interpretability over the expressiveness possible in machine highlights, and we choose a different tradeoff. In TractorBeam, machine annotations are easy to evaluate in context, users are continually reading the associated document content, and additionally iterating on model performance, improving it as they go; we also want users to be able to specify highly complex, interpretive criteria for highlighting. Given these factors, the expressivity of the highlighting is much more a concern for TractorBeam, and that the end behavior of the highlighting model---rather than its reasons for highlighting---are more important, especially in cases where the highlights are not being used to make corpus-level measurement of the incidence of certain kinds of highlight. 

Interpretability, continuous evaluation, and iteration are all hedges around the failure of models, and this does present a real concern. We design TractorBeam to therefore treat machine highlights as inherently less certain than human ones; machine highlights are considered “provisional”, and are not considered part of a user’s schema, until the user explicitly indicates that they are a positive or negative example. We visualize machine highlights as squiggly to convey this sense of uncertainty, relative to a user’s straight-line underlines.

\subsection{Scaffolding Iteration}


In earlier sections, we have discussed TractorBeam’s support for iterative development, e.g. the ability to continuously change its schemata or update machine annotation in-place on schemata. However, as we have noted, LLM behavior is often unpredictable under change. To design around the potential of user changes not achieving their goals, we additionally build features into TractorBeam to allow users to confidently reason about changes that they make.

First, activity in TractorBeam is versioned: schemata are stored using an event sourcing architecture, which keeps a log of replayable events that modify the schema, rather than editing the state itself. This directly enables time-travel debugging on schemata: users can view a log of all changes made to the schema, including new classes, added examples, or edits to descriptions, and roll back to any previous state. Similarly, each instance of a user triggering machine annotation is stored as a “run” that contains metadata about the model configuration that was run, as well as a pointer back to the schema version that was used to construct its system prompt. Each run is also given a shortname in an adverb-verb format, and automatically creates a named bookmark in the schema history, such that every run is linked to a named version of the schema, allowing users to reason about which versions of the schema better captured their intent. All model runs are additionally stored and can be visualized on the document at any point, facilitating some degree of in-place comparison between model runs; a direct, head-to-head comparison between model versions is future work. 

Additionally, TractorBeam provides tools to support rapid iteration: users can choose to apply machine annotations to subsets of document content, such that they can rapidly iterate on e.g. a single page to their satisfaction, before generalizing that annotation out across the whole document. TractorBeam supports annotating individual pages, page ranges, and if the PDF contains a table of contents, annotating sections of documents.

\subsection{Using Highlights}

Finally, TractorBeam provides features to enable navigation both of, and using, both user and machine highlights. Given the potential richness of users’ desired annotation schemata and the possibility of long documents, providing ways to handle large volumes of highlights is necessary. First, we provide a “highlights” tab in the sidebar alongside the “highlighters” view where users edit their schemata, that provides a complete list of all highlights on the document, both user- and machine-originated. This view enables navigation by highlight, in that clicking on a highlight will scroll to it in the document and flash it; it also enables a highlight-centric workflow, in which users can jump from highlight to highlight to quickly skim a document according to their own schema. This highlight view is filterable along several different axes: the highlight’s schema class, its location in the document (page range or document section), and whether it was user- or machine-originated.

TractorBeam also provides a minimap overlay of the scroll bar, visualizing where highlights of different types occur across the document. This presents a previeww of the document similar to the traces of document “wear” presented by \citep{hill_edit_1992}, or the overview panel in \citep{chang_searchlens_2019}. This minimap view aids in not only navigation---users can easily see where in the document certain kinds of annotations occur or cluster, get a preview of the surrounding context in a popover by mousing over the minimap, and navigate to that spot by clicking---but also in users getting a distributional sense of the accuracy of their machine annotations. For example, for an academic reading scientific papers, users can verify that quantitative results and methodology occur largely separately, and that the highlights appear grouped in the document structure as they would expect. The filters from the highlights tab also apply to the minimap, allowing users to distributionally assessing subsets of the highlight types. 

Finally, to enable users to get a more complete overview of their annotations across documents but within a single schema, TractorBeam provides a “Project Manager” view, which provides a CSV-style visualization of annotations within a project across documents. Users can see and filter highlights by schema class, the URL of the PDF on which they were made, the highlighted content, any user notes, and controls to navigate to the highlight in the context of the original document.


\section{Evaluation}
\label{sec:eval}

We ran a formative evaluation of TractorBeam in two modalities: a think-aloud study, and a two-week diary study. In total, we had 13 participants for the think-aloud study, five of whom elected to participate in the diary study. Of these 5, 3 completed the full diary study protocol. Participants were required to be above 18 years of age, to have spent at least 3 years in their current field, and to engage in reading and analyzing PDFs as part of their regular work. All study procedures were approved by our institution’s IRB. 

Our think-aloud study consisted of a single 60-90 minute session with each participant, in which we first demonstrated TractorBeam and its major features. We then had users install TractorBeam on their own computers (though a research computer was made available, were anyone to not want to install software on their computers), and use the tool for the majority of the session on documents that were related to their work. Machine annotations were run using the Anthropic Claude Sonnet 4.6 model as a backend, made available to participants through a researcher-provided API key. We ensured that participants only used documents that were not sensitive, and could be uploaded to a cloud service. Each participant was asked to bring multiple documents. Following participant use of the tool, we concluded with a short debriefing, including asking about demographic information like age, field, and years of experience in their respective fields. We recorded screen shares and audio through all sessions. Following the sessions, we analyzed the recordings using thematic analysis \citep{braun_using_2006}, in which we corrected automatic transcription of recordings, coded the transcripts, and developed themes, which we report below.

All of the participants for the diary study were drawn from the population of participants that had already participated in the think-aloud study, and therefore did not need to be introduced to the tool before beginning the diary study. Participants were asked to use the tool over the course of two weeks, and were compensated \$40 for their time if they completed four diary entries about having used the tool in that time. Participants were also asked but not required to participate in a 30-minute debrief interview after the diary period. We provided continued access to the Anthropic API key through this time, though in the case of P0, their workplace banned the use of Anthropic models midway through the diary period. We provided an OpenAI API key as a replacement. We used AI assistance in generating code for the visualization of diary study activity, and reviewed this code.

Over the course of the diary period, we also collected telemetry data from participants of their use of the tool. This data collection was designed such that no document content or metadata was ever reported; we collected only content generated by the participants themselves: updates to schemata and runs of machine annotation. 

Our participants ranged widely in disciplinary focus: environmental policy scholarship (P0 and P5), materials science (P1, P4, P12), materials informatics (P2), journalism (P3, broadcast and P8, financial), law (P11), and academic research in cultural studies (P10), history (P7), and computer science (P6, P9). Participants had a mean of 10.2 and a median of 7 years of professional experience. Participants were generally in the 25-34 age bracket, with one exception each in the 18-24, 45-54, and 65+ brackets. We report a summary, including how participants self-identified their professional domains, in Table \ref{tab:demographics}

We note that participants approached this tool with widely varying prior experience with both highlighting and AI. While some of our participants commonly highlighted and annotated their reading before using this tool, many found highlighting to be a new modality of document interaction. Similarly, participants had varying levels of both exposure to, and trust in, AI products generally. Several of our participants (P1, P2, P4, P5, P6, P9, and P12) use machine learning approaches in their own research, some use consumer AI products as part of their jobs (P8, P11), and others still largely avoid AI, citing concerns about ethics, trustworthiness, and model capability (P0, P3, P7, P10). These positions were not mutually exclusive: P1, for example, uses AI in research, but generally distrusts the technology.

\begin{table*}[]
    \centering
    \begin{tabular}{cccc}
        \toprule
        Participant ID & Field (self-identification)                  & \multicolumn{1}{l}{Years of Experience} & Age Bracket \\
        \midrule
        P0             & Energy/environmental NGO Scholarship         & 11.5                                    & 25-34       \\
        P1             & Materials Science                            & 8                                       & 25-34       \\
        P2             & Materials Informatics                        & 7                                       & 25-34       \\
        P3             & Broadcast Journalism                         & 10                                      & 25-34       \\
        P4             & Materials Science                            & 45                                      & 65+         \\
        P5             & Environmental Economics                      & 15                                      & 45-54       \\
        P6             & Computer Science/AI                          & 6                                       & 25-34       \\
        P7             & Academic History                             & 5                                       & 25-34       \\
        P8             & Financial Journalism                         & 8                                       & 25-34       \\
        P9             & CL, NLP, Cognitive Science                   & 5                                       & 25-34       \\
        P10            & Asian American Literature, Film, and Culture & 5                                       & 25-34       \\
        P11            & Legal                                        & 4                                       & 25-34       \\
        P12            & Materials Science                            & 3                                       & 18-24       \\
        \bottomrule
    \end{tabular}
    \caption{Participant Demographics}
    \label{tab:demographics}
\end{table*}

\subsection{Findings: Think-Aloud Study}
\label{sec:think_aloud}








\paragraph{Participants found the highlighting affordance was useful in multiple ways.} Participants found machine highlighting to generally be useful in two ways. Some found machine highlights to be useful to guide their reading and attention, especially in conjunction with the use of the minimap, much in the way that \citet{fok_scim_2023} describe how users use Scim, albeit with entirely user-defined categories. Multiple participants noted that in their professional capacity, they were \pquote{hella time constrained}[P3], and spending time in documents was \pquote{a luxury at this point}[P0]. For these participants, the additional structure that TractorBeam provided acted as a personalized index of document content. To hear P1 describe it: \pquote{So that does actually make reading a lot easier...I could see myself highlighting everything automatically and then using that to guide readership as I go. That’s something that I think good writing typically does, kind of, visually guide the reader with headers, et cetera, but… It’s not always the case. So a lot of academic reading is sifting through somebody else’s writing and formatting decisions. So help with that is always appreciated.}

Other participants found highlights useful as a support for their thoroughness in reading papers, and appreciated the call to attention in areas that they had missed. P6 describes having their attention called back to a table that they had missed, because they \pquote{skipped through background}, and P0 describes being reminded of content in a document that they themselves had authored: \pquote{I forgot we talked about the Federal Power Act when we wrote this...that’s helpful to have flagged there.}

TractorBeam’s requirement for structured highlighting encouraged organization and schematization of information, regardless of whether they used machine highlighting.
Participants described the benefits of TractorBeam’s requiring upfront specification of schema classes for individual highlights: \pquote{The strict delineation of categories was helpful...when I’m working on my own, it smooshes together a little bit in my brain}[P7], \pquote{I didn’t find it a drag...it also gives me an opportunity to maybe synthesize what I’m thinking}[P10], \pquote{now that I've described it for an AI model, that has forced me to be more descriptive, and so future me or a collaborator might have a better...understanding of what we were doing in the first place.}[P06]

We also saw support in the think-aloud session for the idea that machine highlights prompted metacognitive behavior from users about their schemata: users frequently used the review of machine highlights as a prompt to create new classes. P6, for example, engaged in a lengthy process of reasoning about why a model may have made a certain annotation, eventually conceding that their schema needed to be updated:\footnote{The names of P6’s schema classes are indicated in \textsc{small caps}.} \pquote{I think it makes sense that this should have been \textsc{creation methodology}, and I think the reason that I had put in \textsc{user information} is because, like, I care about whether demographic data was available as a part of the user information. Yeah. And so maybe then that should have just been a different class.}. This process occurred frequently through the think-aloud sessions. Though users did tend to bring an upfront schematization to documents, those schemata rarely, if ever, remained unchanged when reviewing machine highlights: participants actively split classes, created new ones, and refined definitions as they worked. 
Despite the utility of highlighting, however, multiple users noted the degree of effort it required, especially at startup: \pquote{it was just more work. I was inputting more than I was getting}{P11}. P6, even amid the reconfiguration of their schema, discussed how, while this was causing their schema to better represent their mental model, that \pquote{the goal of this reading is not to create the most robust schema there is}[P6]. That said, both users did feel like improvements in the systems obviated this concern over time.

\paragraph{Participants felt TractorBeam's design choices were trustworthy, and respected their agency} TractorBeam's affordances seemed to address users' existing concerns about AI more broadly. P0 and P1 both mentioned \pquote{real trust issues with LLMs}[P0] and that they are \pquote{always a skeptic about this stuff}[P1]. Participants also discussed their skepticism about particular capabilities of LLMs, especially with regard to questions of socially determined importance or relevance, as in the case of disciplinary standards: \pquote{there is a world in which I might [highlight] "things that look suspicious"...but then I don't know I trust the AI system to do that}[P6], or the degree to which they find LLM systems to be too credulous in their generation: \pquote{I wouldn't base my reading on [AI summaries]...because people just be saying things in their writing}[P1]. 

Participants found that highlights alleviated some of these concerns, for two reasons. First, the presence of highlights on the page allowed the verification of model predictions, through their grounding in page content. Participants frequently sought to establish the context of quotes to understand whether the highlight was correct, e.g. \pquote{the model is right...it’s just that when I saw that highlight for the first time, I had to go looking for the start of the sentence}[P6]. Second, TractorBeam’s design affordances also struck some users as respecting their agency as readers, rather than replacing it, as in the case of P0: \pquote{it feels much more like this tool is oriented towards ‘how can I...support your work?’ whereas orientation of most LLM tools...have this sort of vibe of ‘let me do your work for you.’} P7 goes further: \pquote{having the AI tool be optional is really important for having the tool be something I’m even willing to consider using}[P7]. P10 additionally discussed how TractorBeam’s general utility made them more likely to use the built-in AI.  This was not true of all participants, however, some of whom continued to feel the tension between delegation to an AI and the acceleration it provided. P3, for instance, felt that TractorBeam \pquote{takes out so much thinking that it’s so bad}, and disliked the behavior that the tool encouraged: \pquote{it’s fundamentally not a shortcut, which is good. But also, I find myself looking for things I think of as shortcut moves.} P1, while appreciating machine highlights, made clear that they did not view TractorBeam as appropriate for all types of reading, stating \pquote{if I was reading for a comprehensive phenomenology of what they’re doing, I would need to be much more careful...But for specifically extracting these details, I’m ok with this.}

\paragraph{Participants were able to iterate on machine annotations, and make them more useful over time.} Over the course of sessions, all users reported feeling like they were able to improve the performance of the machine highlighting. This held true regardless of whether the machine highlighter did a good job on first pass; in perhaps the majority of cases, TractorBeam’s initial set of highlights needed heavy correction, both in terms of example instances and the written schema class description updates, before users felt that the machine annotations adequately captured their intent. As P3 puts it, \pquote{I would say it’s the second pass that I saw utility on}. Users generally leaned more on correcting examples than specifying behavior in class names and descriptions, perhaps reflecting that it was \pquote{a lot of effort to put this in actual words}[P6]. Participants could usually arrive at machine annotations that reflected their intent while iterating in their first document; many participants opted organically to begin reading the second document by initiating AI highlighting before starting to read it themselves. Participants appreciated that the highlights reflected their own interests, which one participant characterized as them wanting \pquote{to do different things than a programmer thinks I should want to do.}[P7]. Participants also described being able to gain a sense of how their edits to the schema affected highlighting. P8 described it thus: \pquote{I felt like it improved a lot. I felt like it learned a lot. I felt like I learned how to calibrate the highlights better, which helped me, which helped TractorBeam, which helped TractorBeam help me.}

\paragraph{Specifying how LLMs should behave still presents challenges.} Despite users’ ability to effectively iterate on the functioning of the machine annotations, we nonetheless noticed several indications that behavior specification over time still presents challenges. Perhaps due to the complexity of the tool in the introductory sessions, participants did not use the versioning features; iteration was linear and incremental. Participants also largely did not engage with model configuration unless prompted to by the research team. P10, upon being asked to try a different model to evaluate its performance on some of the more interpretive parts of their schema, stated \pquote{I probably don’t know enough about the different versions...to discern what each one can find better.}[P10].

Users also felt that they did not know what they could ask for from models, typically citing an ignorance of prompting technique. More concerningly, however, we found a frequent tension in what users understood the role of the schema to be. Multiple users did not understand that TractorBeam constructed instructions for machine annotation from their schema descriptions: \pquote{I did not instantly realize that I was telling the model anything. I thought I was telling myself things}[P8]. Users felt a distinction between features that were “for” them versus “for” a model, as in the case of P6, reflecting an uncertainty about what was appropriate to put in a schema class description: \pquote{if I was just writing the project description for myself, I would stop here, but then I’m almost tempted to add instructions}[P6]. Users did not associate changing schema descriptions with shaping machine annotation as much as they did the effects of providing feedback on highlights.

\subsection{Findings: Diary Study}
\label{sec:diary}

\begin{table*}[htbp]
\begin{tabular}{lrrrrrr}
\toprule
Participant & Documents & Events & Highlighted (+/-) & Promoted (+/-) & Class renames & Class re-descriptions \\
\midrule
P0 & 15 & 217 & 24/0 & 65/20 & 3 & 3 \\
P07 & 4 & 64 & 28/0 & 5/1 & 0 & 2 \\
P09 & 11 & 283 & 174/0 & 23/11 & 5 & 3 \\
\bottomrule
\end{tabular}
\caption{Per-participant summary statistics over the study period. Participants generally did not update class names or descriptions, instead favoring adding instances to specify behavior. Two participants accumulated over a hundred annotations in the two-week period.}
\label{tab:diary_stats}
\end{table*}


For the diary study, we collected information about schema updates and model runs from three participants over a two week period. We report general summary statistics about this in Table \ref{tab:diary_stats}. Overall, we observe two modes of interacting with TractorBeam in this data: a style focused on machine highlights, with consistent addition of instances to the schema (P0), and a style much more focused on manual highlights, with machine highlights appearing only after a great deal of manual annotation (P9), or only towards the end (P7). P7 and P9 discussed the utility of the highlighting itself for their reading, indicating that the machine annotations were perceived as optional in the tool, rather than required. P9, for instance spoke about how they \pquote{got what they intended out of highlighting in general}, and how their experience with the tool improved as they learned to \pquote{ignore the machine highlights} as they ran them in the background. P0 also brought up the metacognitive work that this tool encouraged, both at the level of concept definition (\pquote{when I was dealing with sort of bigger, slightly messier concept...it prompted me to sort of, like, think about the boundaries between them, and what would go in what classification, and what the relationship was between them.}), and at the level of how they organize information in their writing (\pquote{I’m definitely thinking about [writing] documents a little bit more in terms of the highlight clusters.}). Participants also continued to feel like TractorBeam’s affordances made AI more trustworthy: P0 discussed not feeling like they were delegating too much to TractorBeam: \pquote{where I felt like TractorBeam was of limited utility because [documents] were well organized...I ended up disregarding it and reading [the document] as is}; P7 felt it was \pquote{a tool premised on language modeling in a way that didn’t feel gross.} 

\begin{figure*}
    \centering
    \includegraphics[width=1\linewidth]{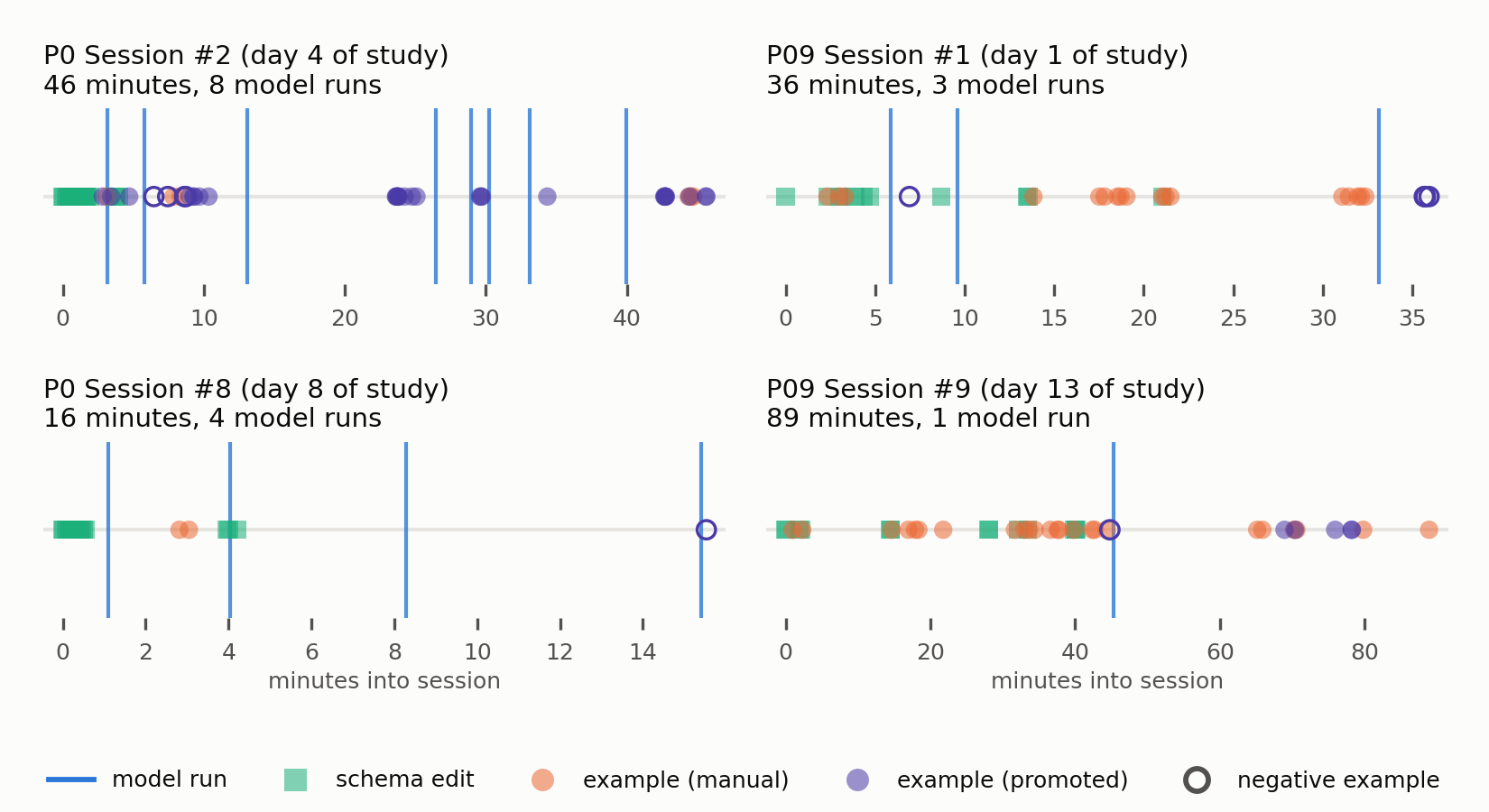}
    \caption{Timeline visualization of user interactions, early and later in the diary study period. Participatns engaged in more schema editing ahead of running machine annotations, and usually fixed machine annotations afterwards. Schema edits in this figure consist of class creation, renaming, and re-description (including initial naming and description). Manual events are highlights from users in the PDF viewers, and promoted examples are machine annotations that received feedback. Note that, due to a bug that caused lag in the storage of model runs, some feedback may appear before the model run was reported. }
    \label{fig:timelines}
\end{figure*}

A common behavior shared by the participants over the period of the diary study was that while participants did successfully iterate on the machine highlight performance over time, they much more frequently did so by adding examples to schema classes than changing the definitions or names of those schema classes. Over the course of the two week period, P0 and P9 created over a hundred examples each to substantiate their classes, but edited the descriptions of those classes fewer than five times. P9 confirmed this in the debrief interview, stating \pquote{I knew that adding a description would probably make the system work better, but I didn’t think it was necessarily worth my time trying to, like, optimize off of that. I kind of just hoped that, implicitly, with more examples, that a language model could pick up on some latent [intent]}. For P9, avoiding descriptions was additionally a way to preserve interpretive ambiguity as they continued to read. In response to a question about whether they were able to stay uncertain: \pquote{I stayed uncertain...a lot of my categories either only had, like, a very brief description, or sometimes no description at all}[P9]. This is corroborated by the data: category creation happened largely towards the beginning of the use period, indicating that users tended to bring categories to their reading, rather than discovering them in exploration during this test; when we asked about this during the debrief interview, participants reported being in more confirmatory reading phases in their work during the time of the diary study. However, we note that category creation did often happen during reading itself: all but one of P7’s created classes (and 4/15 of P9’s) were created directly from the PDF viewer, rather than the sidebar. 
We visualize this in the timeline views in Figure \ref{fig:timelines}. In the timelines, e can see that both P0 and P9 engaged in upfront definition of classes, typically before running machine annotations, and activity following machine annotations consisted almost entirely of adding new instances. In P0's case, this was typically accepting or rejecting machine annotation suggestions; in P9's it was primarily new highlights on the docs, unrelated to machine highlights.

Finally, we observed that participants continued to engage in machine highlighting throughout their use of the tool, while number of instances that they added decreased over time. This trend indicates that participants tended to converge on machine annotation schemata that worked for them over time. This echoes behavior in the think-aloud study, where participants tended to guide their reading of a second document with machine annotations from the beginning, rather than reading it themselves first, as P9 outlines: \pquote{once I had up to 3 or 4 documents in a project I felt more comfortable just running the automatic highlighting from the get-go}. P0 takes this further, arguing that TractorBeam’s primary utility for them was \pquote{for either poorly structured documents, or big documents}, using the minimap as a navigation aid; this was a result of their configuration of the model: \pquote{once I set it up and knew I had a workable schema of highlights, I pretty much let it run and reinforced it as convenient...or correct it.}[P0].  Both P0 and P9 noted, however, the longer-term labor involved in these tools. In P0’s case, they discussed how this could be an effective way to leverage others’ expertise, and having a shared set of knowledge to pull on, envisioning shared highlighters: \pquote{are you trying to navigate the U.S. federal code regarding X?...Then here’s a highlighter that might be helpful to start.} In P9’s case, they discussed how if their schema shifted after reading a variety of documents, the thought of redoing categorization felt \pquote{daunting}.

\section{Discussion and Future Work}




Our observations across both think-aloud and diary study suggest that TractorBeam is a tool that users view as a potentially useful addition to their reading process. In particular, users found TractorBeam’s highlighting affordances---highlights sorted into classes, machine highlights, and the improvement of machine highlights---useful in several ways: as a way to clarify their own thinking, as a skimming aid with which to guide their attention, and as a tool that assists in more comprehensive directed reading. In this section, we further contextualize some of our findings. 

One of our major research questions in this study was to establish whether TractorBeam’s model of collaborative machine annotation was a useful way of reframing LLM outputs, along a few different axes: whether users felt the LLM outputs were trustworthy, whether users felt able to effectively iterate and arrive at a system they felt was useful for reading assistance, and the degree to which the deliberate design frictions that TractorBeam imposed were valuable. On trustworthiness, users felt positively about TractorBeam’s design: they felt that TractorBeam respected their existing agency and reading processes in allowing them to specify the goals of the tool, and in the offer of differential use. Further, the groundedness of annotations in the original document context allowed for a user experience that provided many of the benefits of information extraction or coding---the direction of attention, a better understanding of the structure of documents through where entities occur---while sidestepping many of its concerns. In particular, TractorBeam relaxes the need for quantitative evaluation that looks at exact span matches, in that users were presented with the context, and could focus on parts of that context that may not have been within the original span. \citet{zyto_successful_2012} term this kind of utility “geographic” in discussing how annotation grounded to locations in documents allows for students to index and answer each other’s questions based on content; we find that the geographic nature of annotations serves us here as well. 

Users additionally were able to effectively iterate on their highlighting systems, usually producing systems that suited their intent over the course of 40 or so minutes of use of the tool. This often occurred after the models on first attempt failed in multiple ways. Participants, even those without machine learning knowledge, were able to use both collecting further example instances and re-describing what they were looking for as effective levers to improve model performance  and personalization over time. In favoring primarily collecting examples, participants were additionally able to preserve interpretive ambiguity. In the diary study, use of the machine annotation often stayed steady, despite an abatement in schema iteration, indicating that users asymptotically approach a solution that continues to work for them. We note that users found machine annotations useful despite their black box nature, and sometimes even because of it---annotations that did not fit users’ expectations sometimes caused them to reconsider their schemata in ways that helped them. While TractorBeam’s machine annotations are not interpretable, their presentation as provisional, rather than absolute, combined with the ease of evaluation and iteration, meant that participants did not always need them to be reliable or interpretable in order to find them useful. 

The specification of machine annotations, however, also showed room for future work to better communicate how the machine learning components of our system behaved. In particular, participants did not fully understand what the presented affordances of the machine learning were, or how their descriptions fed into the machine annotation, and correspondingly tended to update models primarily by way of example. This could, however, be related to the act of annotation itself, however: \citet{marshall_exploring_2004} find that annotators of physical books rarely share all of their annotations when digitizing them, instead sharing only a curated subset after further editing. Clarifying what and how the machine annotation relates to personal reading without increasing the complexity and burden of the UI remains a challenge. This, however, is well-studied in prior literature \citet{zamfirescu-pereira_why_2023} study the difficulties of prompting for non-technical users, and \citet{petridis_constitutionmaker_2023} suggest that constructing "constitutions" or prompts from examples might help users specify behavior; we argue that it could also help users formalize their thoughts for unifying principles behind their examples in ways that support sensemaking. This also dovetails with work in NLP that suggests that prompts can express the same range of behavior as large numbers of examples. \citep{lescao_how_2021}.

The design frictions we built into TractorBeam expose further area for future work. P9, in the diary study debrief, discussed an ongoing concern with schema updates over longer time horizons, and P3 in the diary study requested a way to have TractorBeam automatically suggest categories for highlights. Given that multiple participants accumulated over a hundred annotations in the space of a few weeks, the question of how to handle schema changes becomes pressing. PaTAT \citep{gebreegziabher_patat_2023} handles this elegantly: in compositionally constructing codes from rules, it avoids requiring users to manually re-label instances when splitting classes. Earlier work, such as \citet{kulesza_structured_2014}, suggests structuring by labels and tags can facilitate re-labeling diverse concepts as schemata shift. 

The concerns over long-term maintenance additionally echo concerns raised in \citet{karasti_infrastructure_2010}, which discusses the tensions between representations suitable for long-term schemata shared out across organizations. Given that TractorBeam is built around the idea of collaborative annotation, we see it as a reasonable extension of scope to study how to support users collaboratively editing schemata with each other as well. This raises a host of new questions: should users be beholden to synchronize schema representations, and if not, how should divergent changes to schemata be handled? Should AI assistance continue at the level of the individual user, such that users only present AI results to each other if verified, or should AI be involved in teams of multiple humans? 

\section{Conclusion}

We present TractorBeam, a mixed-initiative, browser-based system that supports the construction of personalized, AI-based highlighting systems, framed as collaborative annotation on the content of a paper. We evaluate TractorBeam across a think-aloud study with 13 participants, 3 of whom additionally completed a two-week diary study. We found that participants found the affordance of structured collaborative annotation useful both with and without AI, and that the affordances that it provided made AI more trustworthy for participants who were wary of incorporating AI in their work. Participants were able to iteratively construct model schema that evolved over time, and found our design encouraged them to formalize their thinking process, though they noted ongoing schema management as an area of continued friction. 




\bibliographystyle{ACM-Reference-Format}
\bibliography{main}

\appendix



\end{document}